# On the low-temperature performances of THGEM and THGEM/G-APD multipliers in gaseous and two-phase Xe[*]

A. Bondar[a], A. Buzulutskov[a,†], A. Grebenuk[a], E. Shemyakina[a], A. Sokolov[a], D. Akimov[b], I. Alexandrov[b] and A. Breskin[c]

[a]*Budker Institute of Nuclear Physics, Lavrentiev avenue 11, 630090 Novosibirsk, Russia*
[b]*Institute for Theoretical and Experimental Physics, Bolshaya Cheremushkinskaya 25, 117218 Moscow, Russia*
[c]*Weizmann Institute of Science, 76100 Rehovot, Israel*
*E-mail:* `A.F.Buzulutskov@inp.nsk.su`

ABSTRACT: The performances of THGEM multipliers in two-phase Xe avalanche mode are presented for the first time. Additional results on THGEM operation in gaseous Xe at cryogenic temperatures are provided. Stable operation of a double-THGEM multiplier was demonstrated in two-phase Xe with gains reaching 600. These are compared to existing data, summarized here for two-phase Ar, Kr and Xe avalanche detectors incorporating GEM and THGEM multipliers. The optical readout of THGEMs with Geiger-mode Avalanche Photodiodes (G-APDs) has been investigated in gaseous Xe at cryogenic temperature; avalanche scintillations were recorded in the Near Infrared (NIR) at wavelengths of up to 950 nm. At avalanche charge gain of 350, the double-THGEM/G-APD multiplier yielded 0.07 photoelectrons per initial ionization electron, corresponding to an avalanche scintillation yield of 0.7 NIR photons per avalanche electron over $4\pi$. The results are compared with those of two-phase Ar avalanche detectors. The advantages, limitations and possible applications are discussed.

KEYWORDS: Cryogenic detectors; Two-phase detectors; Geiger-mode APDs; THGEMs; Near Infrared scintillations in noble gases.

---

[*] Work dedicated to the late Boris Dolgoshein, initiator of the two-phase noble-gas detector.
[†] Corresponding author.

# Contents

## 1. Introduction

Over the past decade there has been a growing interest in cryogenic detectors with electron avalanching in Gas Electron Multipliers (GEMs) [1],[2],[3],[4],[5],[6] and Thick Gas Electron Multipliers (THGEMs) [7],[8],[9],[10],[11], operating in noble gas media at low temperatures either in a two-phase [1],[3],[4],[5],[6],[8],[9],[10],[11] or gaseous [1],[2],[7] mode. Such detectors could play an important role in rare-event experiments, such as coherent neutrino-nucleus scattering [12], dark matter search [13], solar neutrino [14] and large scale neutrino [15] detection, and in gamma-ray imaging techniques such as Positron Emission Tomography (PET) [16],[17] and Compton Telescope [18].

    Unlike open-geometry gaseous multipliers, e.g. wire chambers, hole-type multipliers with the avalanche being to a large extent confined within the holes [19], permit attaining high charge gains in "pure" noble gases due to considerably reduced photon-feedback effects. This was demonstrated first in cascaded GEM multipliers operated with high gains in all noble gases at normal and high pressure [20],[21],[22],[23]. Similarly, other hole-type multipliers were shown to operate in pure noble gases at room temperature: Micro-Hole & Strip Plates (MHSPs), THGEMs and Resistive-THGEMs (RETHGEM) in Ne [24],[25], Ar [7],[24],[25],[26],[27], Kr [28] and Xe [7],[24],[25],[27].

    In cryogenic two-phase detectors, the preferable detection media are Ar and Xe. They have the highest cross-sections for nuclear recoils induced by weakly interacting particles, such as neutrino or WIMPs (Weakly Interacting Massive Particles); Kr is excluded due to is natural radioactivity. Xe is also very attractive for medical applications, such as PET and SPECT, due to its higher Z. So far, most promising results have been obtained with two-phase Ar avalanche

detectors; at our laboratory we routinely attained gains of $10^4$ with triple-GEMs [3],[4],[5] and gains of $3\times10^3$ with double-THGEMs [8] irradiated with X-rays and β-particles, stably operating for tens of hours at gains two-fold lower than the maximum. High-gain operation of THGEMs in two-phase Ar was confirmed by other groups, reporting gains reaching 300 [9] and 80 [11] with a single-THGEM, irradiated with X-rays and cosmic rays, respectively (in the latter the originally quoted gain value being adjusted to our gain definition).

The situation with Xe is less clear. At room temperature the maximum gain in Xe reached rather large values, of about $10^4$ in triple-GEM [22] and in double-THGEM [27] multipliers irradiated with X-rays. However in two-phase Xe in saturated vapour, the maximum gains were substantially lower: of about 200 and 150 in triple-GEM [3] and single-GEM [6] multipliers, respectively. On the other hand, little is known about the THGEM performance in Xe at cryogenic temperatures [7] and practically nothing about the operation in the two-phase mode.

In this work, the performances of THGEM multipliers have been studied for the first time in two-phase Xe, as well as in gaseous Xe at cryogenic temperatures. The detector gain, amplitude spectra and noise characteristics were measured both in two-phase and gaseous mode. For convenience, we summarize the existing data on maximum gain attained in two-phase Ar, Kr and Xe avalanche detectors incorporating GEMs and THGEMs.

We also further studied a recently introduced technique [9],[10] of optical readout of THGEM multipliers using Geiger-mode Avalanche Photodiodes (G-APDs or SiPMs [29]), sensitive in either the Vacuum Ultraviolet (VUV) [9] or Near Infrared (NIR) [10]. In detectors requiring ultimate sensitivities, the optical readout might be preferable as compared to charge readout in terms of overall gain and noise. In addition to its high sensitivity in the NIR, the G-APD has a superior performance at cryogenic temperatures, in terms of the gain, noise-rate and pulse-height resolution characteristics, as compared to that at room temperature [30]. In two-phase Ar, a rather high avalanche scintillation yield in the NIR has been observed, of about 4 photons per avalanche electron (over 4π) [10]; it was recently confirmed by direct measurements of primary and secondary NIR scintillations in gaseous and liquid Ar [31]. Since all noble gases are expected to have intense NIR emission [32], in the present work we investigated the optical readout with a THGEM/G-APD multiplier in Xe at cryogenic temperature, comparing the avalanche scintillation yield to that in Ar.

## 2. Experimental setup

The experimental setup was almost identical to that used in our previous measurements with Ar [10]. Accordingly, we consider here mostly features relevant to the performance in Xe. The cryogenic setup is schematically depicted in Fig. 1; the operation principle is shown in Fig. 2. The experimental setup includes a 9 l volume cryogenic chamber of 24 cm internal diameter; it is cooled with liquid nitrogen using a heat-exchanger tube, soldered to a Cu wall of the chamber along its perimeter. The top chamber compartment contained high-voltage (HV) and low-voltage (LV) feedthroughs at room temperature; it was connected to the bottom cryogenic compartment through a bellow joint. The vacuum insulation of the cryogenic chamber was enhanced using a super-insulation made of multilayered aluminized Mylar films. The cryogenic chamber was operated in either a gas-phase or in a two-phase (liquid-gas) mode.

A double-THGEM (2THGEM) assembly was mounted in the cryogenic chamber (see Fig. 1). The THGEMs were similar to that used in [10]; they were made of G10 and had the following geometrical parameters: 25×25 mm² active area, 0.4 mm dielectric thickness, 0.9 mm hole pitch, 0.5 mm diameter holes with 0.1 mm rims around each hole. In a two-phase mode, the detector was operated in equilibrium state, mostly at a saturated vapour pressure of about 1.0 atm, corresponding to a temperature of 165 K. In this mode the chamber was filled with about 0.5 l of liquid Xe (~1.5 kg). The cathode mesh at the bottom was immersed in a ~1 cm thick liquid Xe layer and the double-THGEM was placed within the saturated vapour above the liquid; see Fig. 2. In the cathode gap (the gap between the cathode and the first THGEM) the liquid and gas layer thicknesses were 5 and 5.5 mm respectively.

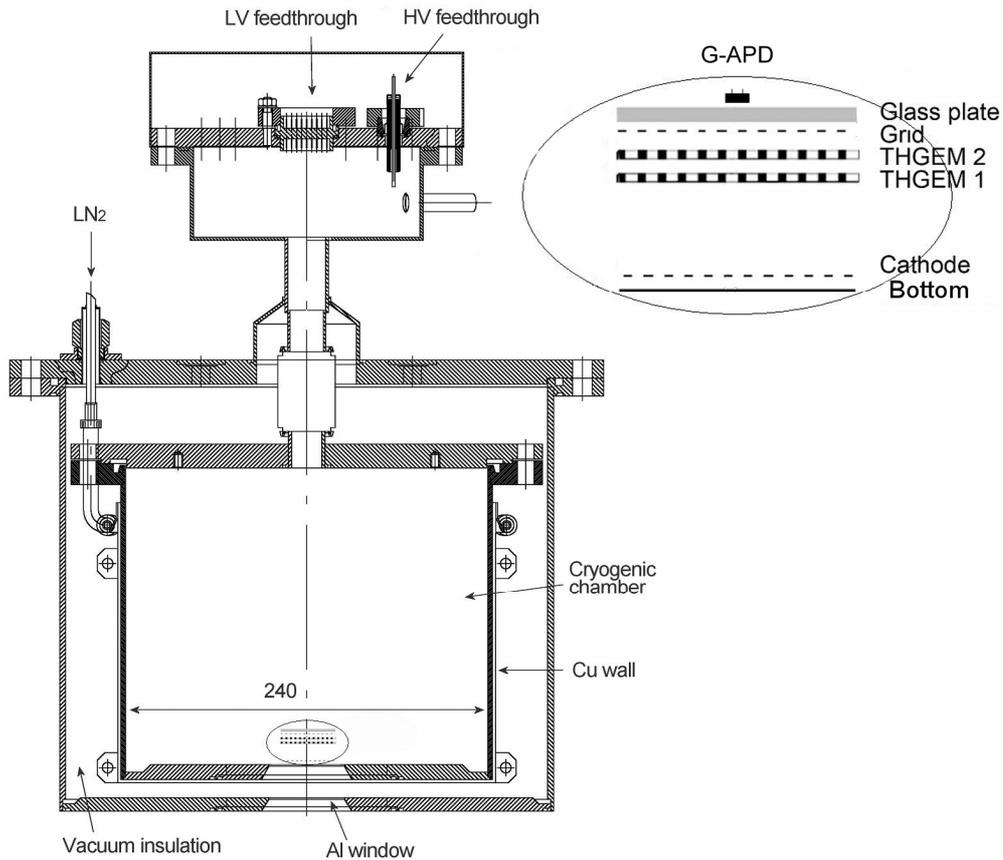

Fig. 1. Design drawing (to scale) of the cryogenic chamber; in the insert is an expanded view of the double-THGEM/G-APD assembly mounted inside.

In the two-phase mode, the liquid Xe level was monitored with an accuracy of 0.5 mm using two methods. Firstly, the cathode gap capacitance, unambiguously related to the liquid layer thickness, was measured during liquefaction procedure, similarly to that of Ref. [5]. Secondly, the liquid layer thickness was calculated from the amount of condensed Xe using the well defined geometry of the chamber bottom (see Fig. 1) and weighing the Xe supply bottle. Both methods gave similar results.

In the gas-phase operation mode the electric field in the cathode gap was kept at ~0.5 kV/cm. In the two-phase mode the electric field across the liquid Xe was kept (in most measurements) at 3.2 kV/cm.

A G-APD (MRS APD "CPTA 149-35", [33]) was placed behind the second THGEM at a distance of 4 mm, separated from the latter by a wire grid and borosilicate glass plate (see Fig. 2). The G-APD was optimized for the green-red range; it had a 4.4 mm$^2$ active area and a

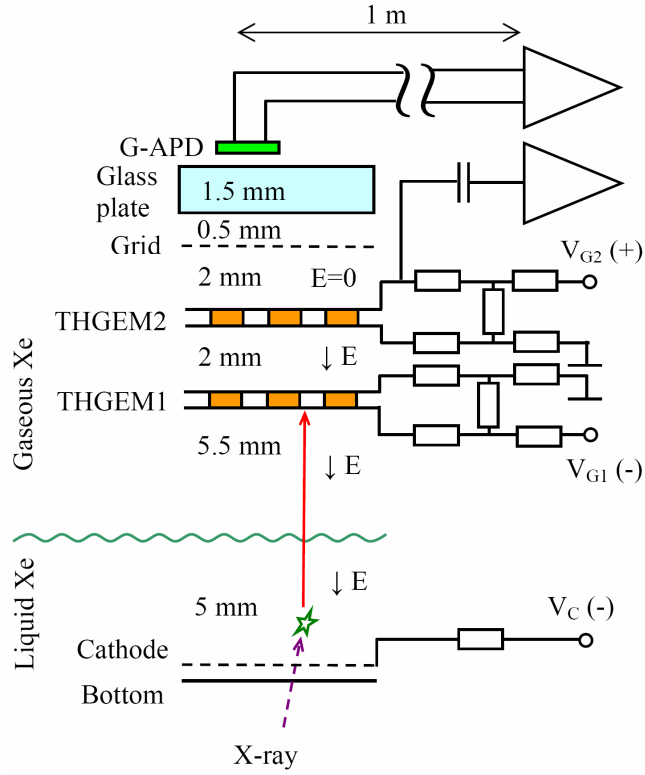

Fig. 2. Schematic view of the experimental setup to study THGEM and THGEM/G-APD multipliers performances in a two-phase Xe avalanche detector (not to scale).

Photon Detection Efficiency (PDE) of 5-10% at the Xe emission range [34]; see Fig. 3. The angle at which it viewed the THGEM active-area (the viewing angle, defined in Ref. [10]), was about ±70º, corresponding to the average solid angle $\Delta\Omega/4\pi=2.7\times10^{-3}$, identical to that of Ref. [10].

 The signals in the detector were induced by radiation entering the cryogenic chamber through Al windows at the chamber bottom: 15-40 keV X-rays from a pulsed X-ray tube with a Mo anode operated at a voltage of 40 kV, X-rays from a $^{241}$Am source with a dominant line at 60 keV and 511 and 1275 keV gamma-rays from a $^{22}$Na source. The charge gain of the THGEM multiplier was measured with pulsed X-rays with an amplifier shaping time of 10 μs, similar to that in our previous works [10]: the gain was defined as the ratio of the output charge to the input charge of the gas multiplier. In other words, it is equal to the pulse-height of the avalanche (anode) signal of the THGEM multiplier divided by that of the calibration signal. The latter was recorded at the first electrode of the first THGEM in special calibration runs, when the cathode gap was operated in an ionization collection mode, the first THGEM acting as an anode (i.e. with no gain). The gain-voltage characteristic obtained in this way allowed us to asses the number of initial electrons (prior to multiplication) when operated with $^{241}$Am and $^{22}$Na sources: it was just the ratio of the output signal charge of the THGEM multiplier to its gain.

 The experimental setup was not optimized for operation with very-high purity liquid Xe; the cryogenic chamber contained the G10-made THGEMs and other support plates, Kaprolon (Polyimide-6) piers and fittings, HV-protective Mylar films, Kaprolon HV connectors with epoxy insulation, ceramic-insulated HV connectors, glass-insulated LV connectors, Teflon tubes and Teflon-coated wires. In an effort to improve the purity, the experimental chamber with the THGEM assembly was heated to 70 ºC and pumped to a vacuum of $10^{-5}$ Torr for several days. Also, the Xe gas was initially purified from electronegative, organic and nitrogen impurities using a NiCr catalyst and molecular sieve purification system of the liquid-Xe calorimeter of the CMD detector [35]; this provided an electron lifetime of 15 μs in liquid Xe, measured with a small test chamber. However in the cryogenic chamber, the liquid Xe purity was degraded, in spite of the additional oxygen and water removal by an Oxisorb filter [36] during the cooling procedure. The degradation was presumably caused by outgasing from the substantial amount of low-purity materials present in the cryogenic chamber, as described above. The liquid Xe purity was monitored measuring the charge extracted from the liquid to the gas phase, namely measuring the calibration signal amplitude induced by pulsed X-rays and that of the 60 keV X-rays from $^{241}$Am source, as discussed in Section 4.3. During the major measurement runs the purity level was stable: the electron life-time in liquid Xe was 1.2 μs, corresponding to an oxygen-equivalent impurity content of $5\times10^{-7}$ [37].

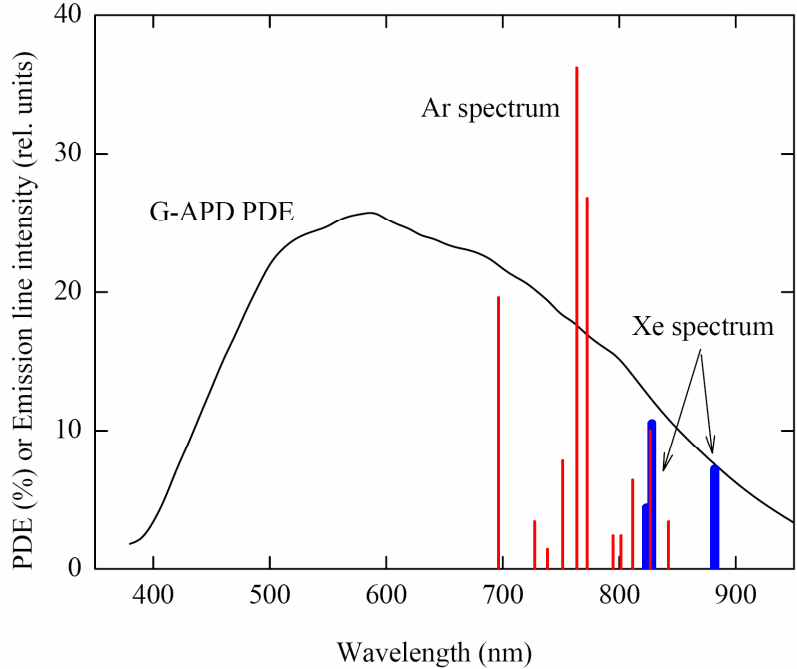

Fig. 3. Scintillation emission spectra of Xe (heavy blue lines) and Ar (light red lines) gases within the G-APD sensitivity range, taken from [32], and the photon detection efficiency (PDE) spectrum of the G-APD (MRS APD "CPTA 149-35") as provided by the producer [34]. The relative intensity of Ar and Xe emission lines corresponds to the relative avalanche scintillation yield measured in this work and in [10].

This purity level permitted to thoroughly study the avalanche characteristics of our gaseous and two-phase Xe avalanche detectors, as well as those of avalanche scintillations in the NIR: such low impurity content has negligible effect on both the charge gain and the light yield in the NIR in noble gases. Indeed, even for Penning mechanism of electron avalanching in noble gases, which is known to be most sensitive to impurities, the impurities start affecting the charge gain characteristics from the content of the order of $10^{-5}$ (see for example [2]), which is considerably higher than that of the present work. Regarding NIR avalanche scintillations in noble gases, it was shown that these are much less sensitive to impurity content, as compared to VUV scintillations [38]: in particular, the addition of up to 2% of $N_2$ to Ar did not affect the Ar emission in the NIR; the same behaviour is expected for Xe.

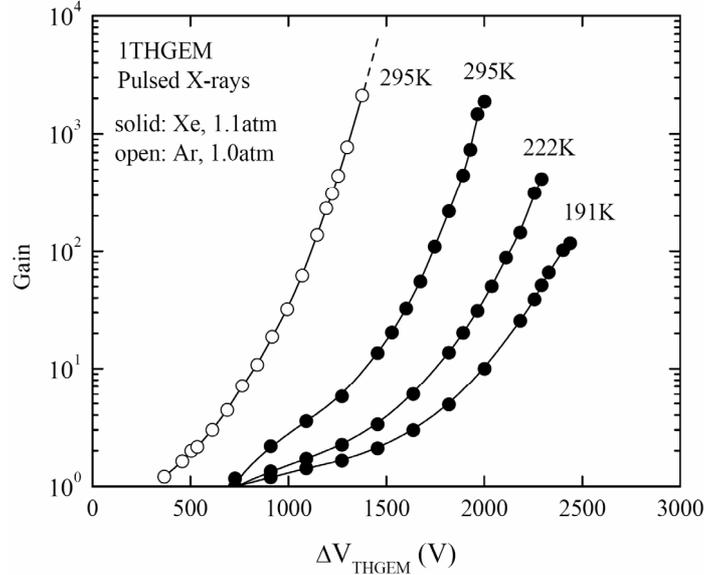

Fig. 4. Gain-voltage characteristics of a single-THGHEM (1THGEM) multiplier in gaseous Xe at different temperatures and a pressure of 1.1 atm. For comparison that in gaseous Ar at room temperature and 1 atm [8] is shown. In Xe the maximum gains were limited by discharges, while in Ar the discharge limit was not reached.

    Other details of the experimental setup and procedures are presented elsewhere [10].

## 3. THGEM multiplier performance in gaseous Xe

Figs. 4 and 5 illustrate THGEM multipliers performances in gaseous Xe. Fig. 4 shows gain characteristics of a single-THGEM multiplier in gaseous Xe at room and cryogenic temperatures at a fixed pressure of 1.1 atm, in comparison with that of Ar at room temperature [8]. Unlike in Ar, here the maximum gains in Xe were limited by discharges. Similarly to [27], the operation voltages in Xe at room temperature were substantially larger than those in Ar, reflecting the fact that the first Townsend coefficient in Xe is smaller than that of Ar [39]. At room temperature the maximum charge gain in Xe reached a value of ~2000, rapidly dropping with the temperature decrease; it is obviously due to an increase of the gas density, similarly to that observed in high pressure Xe for triple-GEM [22] and THGEM [27] multipliers. At 191 K and 1.1 atm, at a voltage across a THGEM of 2500 V, the maximum gain amounted at 100. Notice that these conditions roughly correspond to that of saturated vapour in a two-phase Xe system at 165 K and 1.0 atm, in terms of gas density. That means that from this point of view, the THGEM multiplier should be able to operate in two-phase Xe at appreciable gains, as will be confirmed in the next sections.

    This statement was also confirmed for the double-THGEM multiplier in gaseous Xe: Fig. 5 shows its gain characteristic at 178 K and 1 atm, i.e. at conditions close to those of saturated vapour in two-phase Xe. Here the maximum gain was measured to be about 600 at a voltage of 2100 V; it was limited by the onset of intense noise. For comparison, the other gain characteristics at room temperature are shown: that of the double-THGEM in Xe, that of the

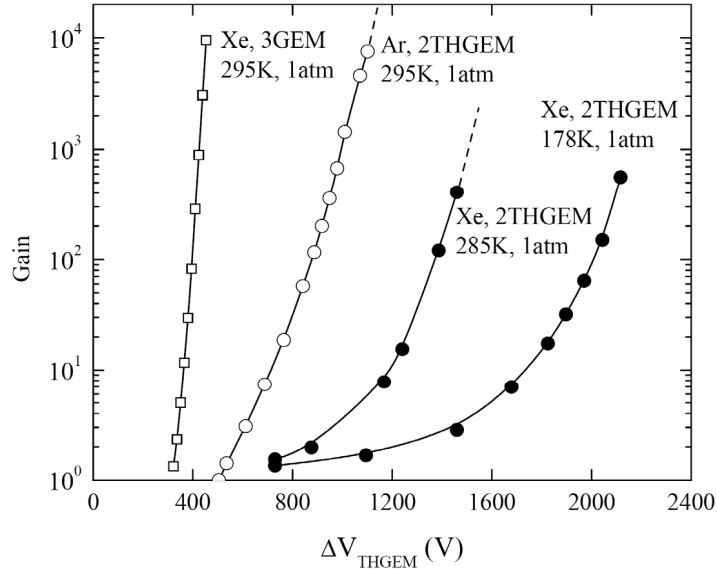

Fig. 5. Gain-voltage characteristics of a double-THGHEM (2THGEM) multiplier in gaseous Xe at 285 and 178 K, at a pressure of 1.0 atm. For comparison that of the double-THGEM in gaseous Ar [8] and that of the triple-GEM (3GEM) in gaseous Xe [22] are shown at room temperature. In Xe at 178 K and at 295 K the maximum gains were limited by intense noise or discharges for the 2THGEM and 3GEM respectively. In other configurations the maximum gains were not reached.

double-THGEM in Ar [8] and that of the triple-GEM in Xe [22]. At room temperature, the operation voltages in Xe of the double-THGEM were considerably larger than those of the triple-GEM; these were also larger than those in Ar, similarly to that observed with the single-THGEM multiplier.

The pulse-height distribution of X-rays from a $^{241}$Am source, measured with the double-THGEM multiplier in gaseous Xe at cryogenic temperature is illustrated in Fig. 6; the conditions were: T=200 K, p=0.73 atm, double-THGEM gain of 350. The amplitude is depicted in number of initial electrons (prior to multiplication). One can distinctly see a dominant 60 keV X-ray photoelectric peak, as well as broad distribution at smaller amplitudes due to lower-energy X-ray lines produced by daughter isotope $^{237}$Np, grouped near 14, 18, 21 and 26 keV [40]. This is in contrast to measurements in Ar where the 60 keV peak was not seen due to photoelectron escaping from the detection area. The peak charge, of about 2500±300 e, turned out to be close to the expected value of 2727 e, using W=22 eV/electron-ion pair [41]. This indicates upon negligible electron losses in gaseous Xe due to recombination and attachment.

## 4. THGEM multiplier performance in two-phase Xe

### 4.1 Hole-type multipliers performances in two-phase Ar and Xe: previous results

It should be remarked that the performance of hole-type multipliers in two-phase Ar and Xe is not fully understood: not all multiplier types were able to operate with electron multiplication in saturated vapour. In two-phase Ar, while G10-based THGEM multipliers successfully operated for tens of hours with gains reaching several thousands [8], others, with resistive electrodes (RETHGEM, [26]) did not show multiplication in an equilibrium state [8]. Though the geometrical parameters of the RETHGEMs were somewhat different from the regular THGEMs, namely smaller diameter holes without protective rims, this can hardly explain the considerable difference in multiplication. Note that Kevlar-based THGEM which also yielded high gains in two-phase Ar [8], had the same hole diameter as the RETHGEM; also, the rimless G10-based THGEM of Ref. [9] effectively operated in two-phase Ar with significant gains. It could be that the specific properties of the hole and electrode surfaces, i.e. the wetting capability, might play a role, as well as the electrode's heat conductivity.

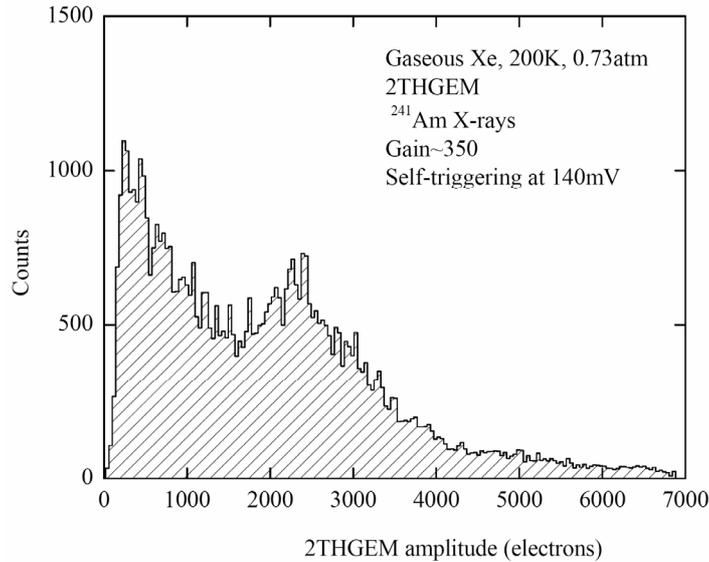

Fig. 6. Amplitude distribution in a 2THGEM multiplier operated in gaseous Xe at 200 K and 0.73 atm, induced by $^{241}$Am X-rays; charge gain: 350. The amplitude is expressed in number of initial electrons (prior to multiplication).

As concerns Xe, the problem is also knotty. So far, only two groups have reported on GEM multipliers performances in two-phase Xe with some multiplication [3],[6], namely with moderate gains not exceeding 200. On the other hand, in Ref. [42] it was reported on unsuccessful performance of the Micromegas multiplier in two-phase Xe: the multiplication was not stable; in half an hour it collapsed due to condensation of Xe within the Micromegas mesh. As was supposed by the authors, this was caused by Xe-vapour migration to regions of large electric-field non-uniformity due to the high polarisability of Xe atoms.

In this respect, the present successful results reported below may be of high significance.

### 4.2 Gain and electron emission characteristics

We have experienced here a successful operation of a G10 THGEM multiplier in two-phase Xe: stability over several hours at relatively high gain, with good reproducibility in different experimental runs. Fig. 7 shows gain characteristics of the double-THGEM multiplier in two-phase Xe, at 165 K and 1 atm, in comparison with that of the triple-GEM obtained earlier [3]. The double-THGEM reached a maximum gain of 600, namely three-fold higher than the triple-GEM that operated at lower voltage. On the other hand, it is five-fold lower than the maximum double-THGEM gain obtained in two-phase Ar [8]. Moreover, in contrast to Ref. [42], there was no indication on eventual Xe-vapour condensation that would affect the THGEM multiplier performance in two-phase Xe.

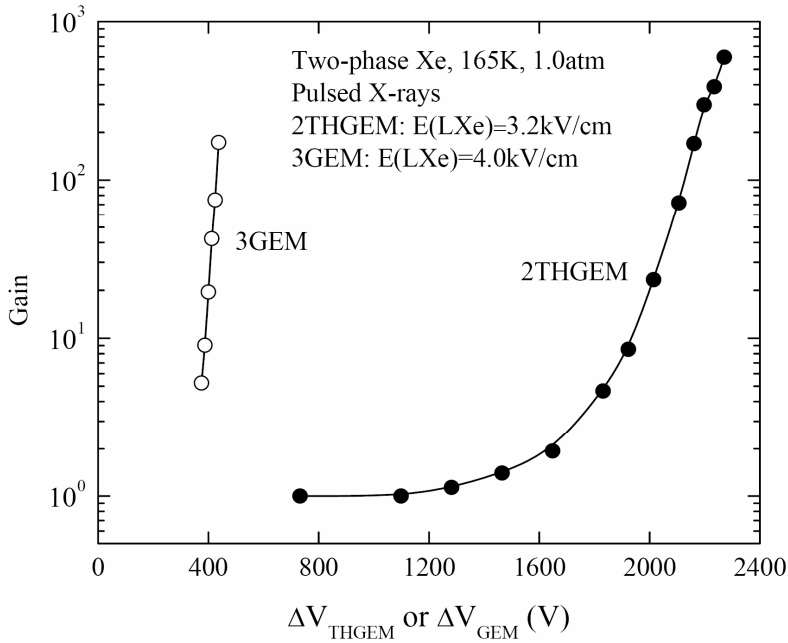

Fig. 7. Gain-voltage characteristic of a double-THGEM multiplier in two-phase Xe at 165 K and 1.0 atm. For comparison that of a triple-GEM [3] is shown. The maximum gains were limited by discharges.

Fig. 8 illustrates the reproducibility of gain characteristics in two-phase Xe, reflected by results obtained in two measurements separated by two weeks. The gain characteristic in gaseous Xe, at a gas density close to that of saturated vapour in two-phase Xe, is shown for comparison. From this comparison, one may conclude that the maximum gain, as well as the operation voltage, are determined essentially by the Xe gas density. If there was an eventual Xe vapour condensation, it had no influence on the THGEM multiplier performance in two-phase Xe.

Fig. 9 characterizes electron emission properties of a two-phase Xe system: the anode signal amplitude from the 2THGEM multiplier induced by pulsed X-rays is shown as a function of the electric field in the liquid and gas phases, at a gain of about 50. One can see that the electron emission through the liquid-gas interface started at some threshold field, of about 1.6 kV/cm in the liquid; the electron emission efficiency plateau was not reached even at field-values of 4 kV/cm, in accordance to prior observations [1],[43]. Note the very large electric field values applied in the gas phase, reaching 7.5 kV/cm (presumably, this field was indeed somewhat larger, since its known extension from the THGEM holes was not taken into account in the calculations). This value is similar to that applied in dark matter search experiments with two-phase Xe detectors [44], providing necessary conditions for reasonable electron emission through the liquid-gas interface (with emission efficiency exceeding 80% [43]), on the one hand, and intense electroluminescence signal in the gas phase, on the other hand.

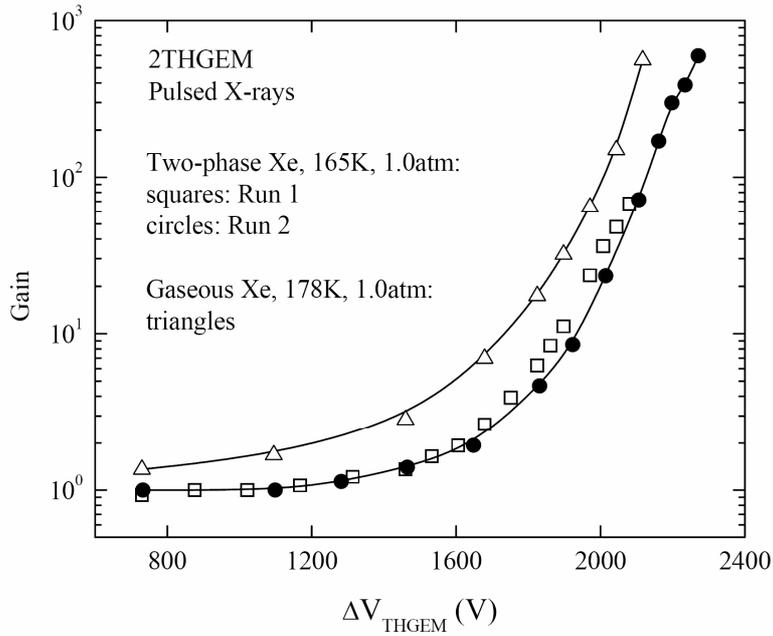

Fig. 8. Reproducibility of gain-voltage characteristics of a 2THGEM multiplier in two-phase Xe at 165 K and 1.0 atm; the two measurements were separated by two weeks. A gain curve in gaseous Xe at 178 K and 1 atm is shown for comparison. In two-phase Xe in Run 2 and in gaseous Xe the maximum gains were limited by discharges and intense noise respectively. In Run 1 the maximum gain was not limited by discharges.

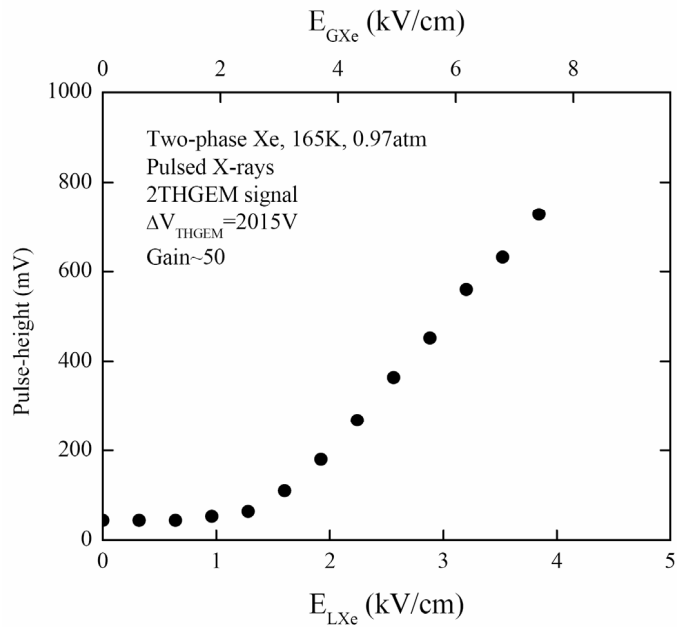

Fig. 9. Pulse-height of the anode signal from the 2THGEM multiplier in two-phase Xe as a function of the electric field in the liquid (bottom scale) and in the gas phase (top scale), at a 2THGEM gain of about 50. In the electric field calculation, the field extension from the THGEM holes was not taken into account.

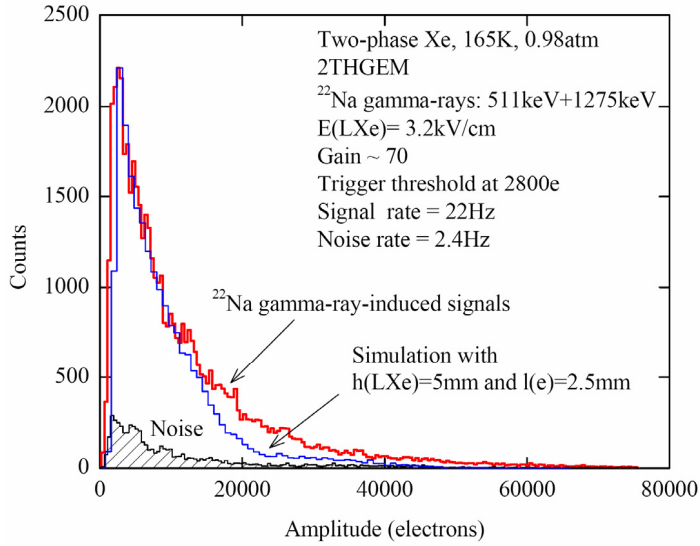

Fig. 10. Amplitude distribution from the 2THGEM multiplier in two-phase Xe, induced by 511 and 1275 keV gamma-rays from $^{22}$Na source converted in the liquid, at 2THGEM gain of 70 and charge amplifier shaping time of 10 μs. The noise contribution is also shown. For comparison, the distribution simulation is presented for liquid layer thickness h(LXe)=5 mm and electron drift attenuation length in the liquid l(e)=2.5 mm. The amplitude is expressed in initial number of electrons (prior to multiplication).

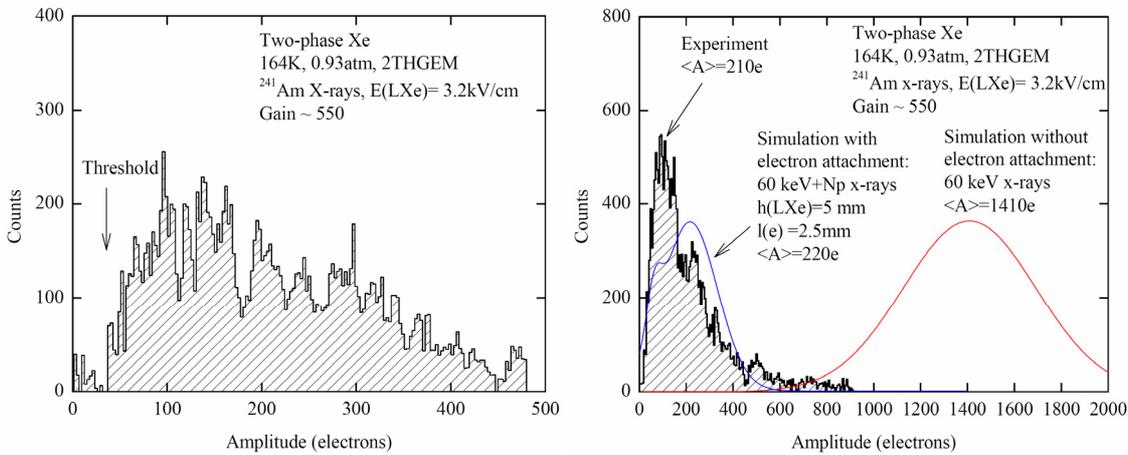

Fig. 11. Amplitude distributions from the 2THGEM multiplier in two-phase Xe, induced by $^{241}$Am X-rays converted in the liquid: two measurement runs are presented with finer (left) and coarser (right) amplitude bins. 2THGEM gain: 550; charge amplifier shaping time: 10 μs; noise contribution is subtracted. For comparison, the distribution simulations are shown: for 60 keV X-ray line without electron attachment and for 60 keV line along with softer Np X-ray lines with electron attachment in the liquid, with liquid layer thickness h(LXe)=5 mm and electron drift attenuation length l(e)=2.5 mm. The amplitude is expressed in initial number of electrons (prior to multiplication).

### 4.3 Amplitude and noise characteristics

In the present work the electron attenuation drift length in the liquid (2.5 mm; see below) was smaller than the liquid-layer thickness (5 mm), resulting in a dependence of the detected signal amplitude on the gamma-ray conversion depth in the liquid. Therefore, the 511 and 1275 keV $^{22}$Na gamma peaks were not resolved, having their conversion points almost uniformly distributed across the liquid depth. This is seen in Fig. 10 showing the amplitude distribution from the 2THGEM multiplier in two-phase Xe induced by 511 and 1275 keV gamma-rays converted in the liquid. The shape of the distribution and its absolute amplitudes were well reproduced by Monte-Carlo simulation, taking into account the following effects: the gamma-ray absorption in the liquid, ionization charge recombination in the liquid, electron emission from the liquid and electron attachment in the liquid, the latter taken with the electron attenuation drift length of 2.5 mm (Fig. 10). Nevertheless, even at a moderate gain of ~70, the gamma-ray events were effectively counted by the detector at a detection threshold of 260 mV corresponding to about 2800 initial (prior to multiplication) electrons; their counting rate was in agreement with the expected one, significantly exceeding the 2 Hz noise rate (see Fig. 10). Accordingly, one may conclude that the THGEM-based two-phase Xe avalanche detector is capable to count efficiently 511 keV gamma-rays at moderate gains, of the order of 100, which is of relevance e.g. for PET applications.

The absorption length of the 60 keV $^{241}$Am X-rays in liquid Xe is only ~0.4 mm; for accompanying softer X-rays it is even smaller. This resulted in a rather thin X-ray conversion region in the vicinity of the cathode. The latter in turn permitted to observe a broad peak in the amplitude distribution at the detector's maximum gains, at a threshold corresponding to about 30 initial electrons: see Fig. 11 (left). Note that at this rather low threshold the noise rate was only ~6 Hz. The average of the distribution, after subtracting the noise contribution, amounted at 210 initial electrons (prior to multiplication): see Fig.11 (right). This value should be compared to the estimated number of 1410 initial electrons, in the absence of electron attachment but taking into account recombination and electron-emission efficiency at a given electric field within the liquid: see the appropriate simulated distribution in Fig. 11 (right). To estimate once more the electron attenuation drift length in the liquid, this distribution average was reproduced by Monte-Carlo simulation, taking into account as before the effects of X-ray absorption, charge recombination, electron emission and electron attachment in the liquid, as well as the X-ray absorption in Al windows at the chamber bottom. In the simulation, all groups of X-ray lines were considered, namely that of 60 keV and that of softer X-rays from a daughter isotope $^{237}$Np. The best agreement between the measured and simulated values was attained for the electron attenuation drift length of 2.5 mm (see Fig. 11 (right)), in accordance with the gamma-ray irradiation data. It corresponds to the electron life-time in liquid Xe of 1.2 μs, as was already mentioned in Section 2.

**4.4 Summary of GEM and THGEM multipliers performances in two-phase Ar, Kr and Xe**

The present work concludes our series of gain measurements of GEM and THGEM multipliers in two-phase noble-gas avalanche detectors [3],[4],[5],[8],[45]. We found it useful to summarize in Table 1 all presently existing data on maximum gains attained in two-phase Ar, Kr and Xe avalanche detectors incorporating these hole-multipliers, irradiated with X-rays or with MIPs (minimum ionizing particles). The following conclusions can be drawn from Table 1:

1. In a sequence "Ar, Kr, Xe" the maximum gain of two-phase avalanche detectors decreased from Ar to Xe by more than an order of magnitude and by half an order of magnitude for GEM and THGEM multipliers respectively.
2. In two-phase Xe the THGEM multipliers permitted reaching higher gains than GEMs: of the order of 500.
3. In terms of the maximum reachable gain in two-phase avalanche detectors, the most efficient were Ar-operated ones: the maximum gain reached values of several thousands, both in GEM and THGEM multipliers.
4. The data obtained by different groups on the maximum gain of single-THGEM multipliers in two-phase Ar are in good agreement. It should be remarked that here the original gain value of 30 reported in Ref. [11] was normalized according to our gain definition, namely to 80, taking into account the reduced electron transparency of the extraction grid at given electric fields of Ref. [11].

| Group | Two-phase medium | Multiplier type | Active area, cm | Typical maximum gain | Reference |
|---|---|---|---|---|---|
| BINP | Ar | 3GEM | 2.8×2.8 | $(5-10) \times 10^3$ | [3],[5] |
| BINP | Ar | 2THGEM | 2.5×2.5 | 3000 | [8] |
| BINP | Ar | 1THGEM | 2.5×2.5 | >200 | [8] |
| Sheffield | Ar | 1THGEM | 4×4 | 300 | [9] |
| ETH Zurich | Ar | 1THGEM | 10×10 | 80 | [11] |
| BINP | Kr | 3GEM | 2.8×2.8 | 400 | [45] |
| BINP | Xe | 3GEM | 2.8×2.8 | 200 | [3] |
| Coimbra | Xe | 1GEM | 2.8×2.8 | 150 | [6] |
| BINP | Xe | 2THGEM | 2.5×2.5 | 600 | Present work |

Table 1. Summary of maximum charge gains reached with GEM and THGEM multipliers operated in two-phase Ar, Kr and Xe, obtained by different groups, irradiated with X-rays and MIPs. The gain is defined as that in the present work (see section 2).

Our general conclusion is that the maximum gains achieved in two-phase avalanche detectors, namely several thousands in two-phase Ar and half a thousand in two-phase Xe, might not be sufficient for efficient single-electron counting, recording avalanche-charge in self-triggering mode. Accordingly, ways of increasing the overall gain (or detected signal) should be looked for. A possible solution investigated is the optical readout of THGEM avalanches using G-APDs; it is discussed in the following section.

## 5. THGEM/G-APD multiplier performance in gaseous Xe

The optical readout of THGEM multipliers, recording avalanche-induced photons by G-APDs [9],[10], would substantially increase the overall gain; the latter being the number of THGEM-emitted photons recorded by the G-APD multiplied by its amplification factor. This could provide an effective single-electron counting in two-phase avalanche detectors. In addition, a multi-channel G-APD readout with overlapping fields-of-view and coincidence between channels, would effectively suppress single-channel noise.

Following our recent results of THGEM/G-APD readout in two-phase Ar [10], we have investigated this concept in gaseous Xe at cryogenic temperature. It is essential to underline that in both cases G-APDs without Wavelength Shifter (WLS) coating, i.e. insensitive to the VUV emission of Ar and Xe, were used. It is also important to note that in high-pressure noble gases the VUV emission, caused by excimer radiative decays, was generally believed to dominate over atomic emission in the visible and infrared regions [41].

However, 20 years ago it was suggested that this statement might not be valid due to the discovery of intense atomic emission scintillations in the NIR in practically all noble gases [32]: in Ar it extended over the wavelength range of 690-850 nm [32] and in Xe over 800-1600 nm [32],[46]. Fig. 3 shows emission spectra of Ar and Xe in the sensitivity range of the G-APD, i.e. at wavelengths of 400-950 nm; for Xe, the spectrum consisted of 3 atomic emission lines, namely of Xe I at 823.2, 828.0 and 881.9 nm [32]. The emission lines are produced by transitions between the atomic states of the Xe ($5p^5$ $6p$) and Xe ($5p^5$ $6s$) configurations, for Xe, and Ar ($3s^2 3p^5$ $4p$) and Ar ($3s^2 3p^5$ $4s$) configurations, for Ar [47]. One may conclude from this figure that the photoelectric yield of the THGEM/G-APD multiplier should be substantially lower in Xe compared to that of Ar, since the major part of the Xe emission spectrum, with a maximum at 1300 nm [46], resides out of the G-APD sensitivity range.

The NIR scintillation yield has been recently measured in gaseous Ar [31]: for primary scintillations it turned out to be rather high, amounting to $1.7 \times 10^4$ photon/MeV or 0.44 photon/e. This is comparable with the lower limit of primary NIR scintillation yield established formerly for gaseous Xe [48]: $\geq 21*10^3$ photon/MeV.

In this work we were interested primarily in secondary avalanche-induced NIR scintillations, since only these ones are useful for the optical readout. Note that there are two kinds of secondary scintillations: those in a proportional scintillation mode (i.e. prior to avalanche multiplication) and those in an avalanche-scintillation mode; the light yields in these might be different. While there are no data on secondary NIR scintillations in gaseous Xe, there exist such data in Ar. In particular, the avalanche-scintillation yield was estimated to be 4-5 photons per avalanche electron in a THGEM operated in two-phase Ar at 87 K [10]. These data will be used in the following for comparison with those in gaseous Xe. It should be remarked that avalanche scintillation yields in both gaseous Xe and two-phase Ar can be compared since in both operation modes light emission occurs in the gas phase.

The present measurements with the THGEM/G-APD multiplier were performed in gaseous Xe at 200 K and 0.73 atm (gas density corresponding to that at 1 atm and room temperature), the measurement procedure being identical to that in [10] for Ar. The characteristics of the G-APD at these conditions are illustrated in Fig. 12, depicting G-APD noise signals and their pulse-height distribution. The pulse-height was measured by integrating the area under the unipolar pulses. One can see that single-, double- and triple-pixel signals are well resolved, demonstrating the effective G-APD performance in single-photoelectron counting mode at 200 K.

Avalanche-induced scintillations occurring in the holes of the second THGEM were detected with a bare G-APD (without WLS, i.e. insensitive to VUV), the scintillation being most probably in the NIR as discussed above. The signals were induced by [241]Am X-rays. Typical avalanche-scintillation and avalanche-charge signals of the double-THGEM, operated at a charge gain of 350, are presented in Fig. 13; one can see that the scintillation signals consist of a number of fast pulses, distributed over the time structure of the charge signal.

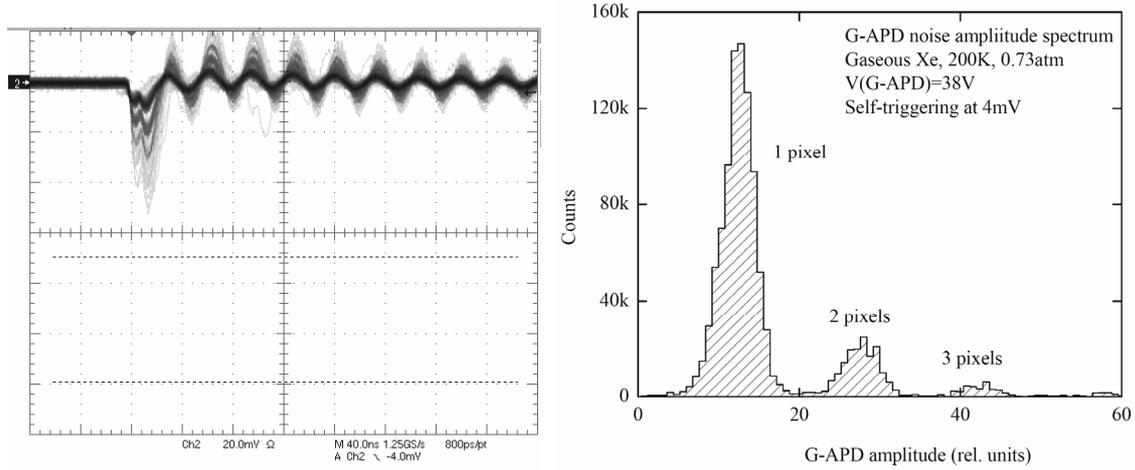

Fig. 12. G-APD noise signals (left) and noise pulse-height spectrum (right) in gaseous Xe at 200 K at a bias voltage of 38 V. The pulse-height was derived from the area under the G-APD unipolar pulses, integrated over 400 ns.

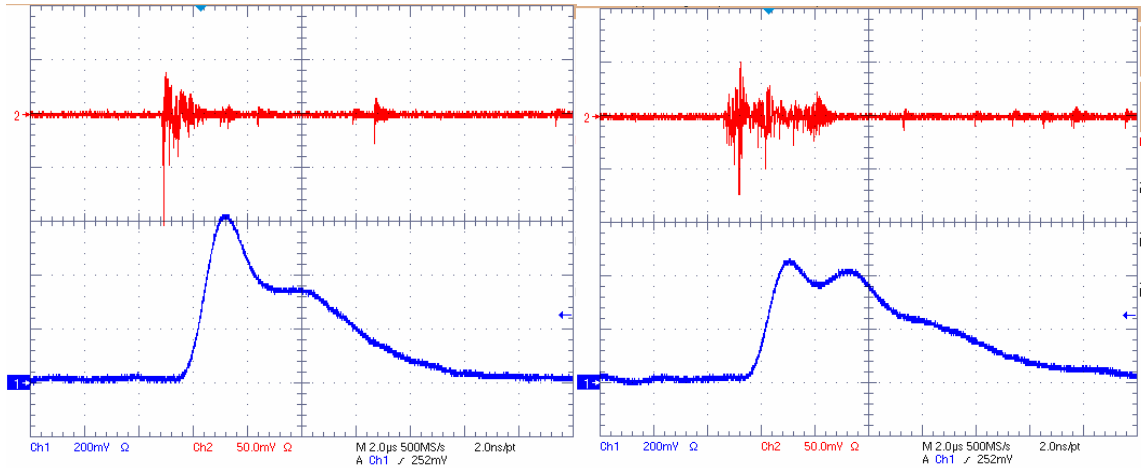

Fig. 13. Typical G-APD avalanche scintillation signals (upper traces) and 2THGEM avalanche-charge signals (lower traces) in gaseous Xe at 200 K and 0.73 atm, induced by [241]Am X-rays. 2THGEM gain: 350; G-APD bias voltage: 38 V; charge-amplifier shaping time: 0.5 μs. The horizontal scale is 2 μs/div. The vertical scale for the scintillation signal is 50 mV/div.

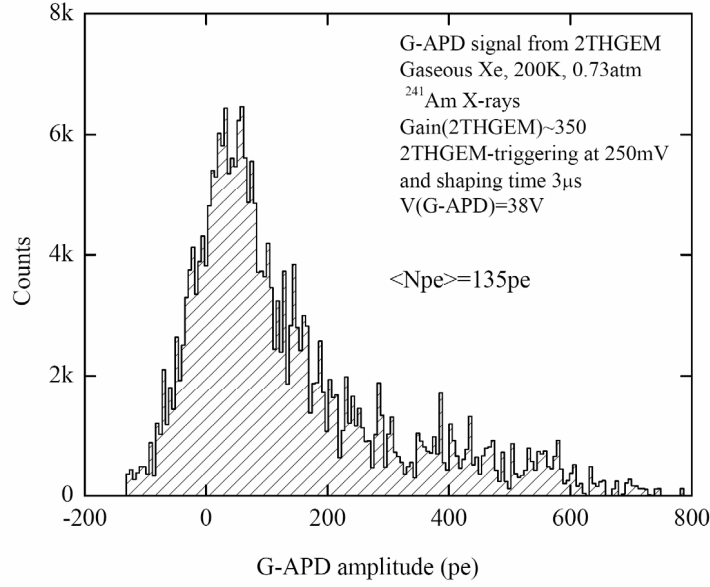

Fig. 14. Avalanche scintillation pulse-height spectrum from the 2THGEM read out by the G-APD, in gaseous Xe at 200 K and 0.73 atm. The amplitude is expressed in number of photoelectrons (pe) derived from the G-APD signals (Fig. 12) integrated over 20 μs, not corrected for nonlinearity and cross-talk. The signals were induced by $^{241}$Am X-rays. 2THGEM gain: 350; G-APD bias voltage: 38 V. The average amplitude is 135 pe.

Fig. 14 characterizes the combined THGEM/G-APD multiplier yield: the avalanche-scintillation amplitude distribution in Xe is shown at a double-THGEM gain of 350; the G-APD signals were integrated over 20 μs. The raw average value of the distribution (135 photoelectrons) was corrected for the G-APD nonlinearity and cross-talk (the latter being deduced from Fig. 12), as discussed in [10]. This resulted in an absolute yield of scintillation-induced 120 photoelectrons per $^{241}$Am X-ray absorbed in the Xe gas. Each X-ray produces in the gas on the average about 1700 initial electrons, the latter being measured by averaging the charge amplitude spectrum (similar to the one shown in Fig. 6).

Consequently, at the present THGEM gain and solid angle, under which the G-APD was viewed by the THGEM holes (it was identical to that of Ref. [10]), the combined THGEM/G-APD multiplier yielded in gaseous Xe an average number of avalanche-induced photoelectrons per *initial electron* of:

$$Y_{THGEM/G-APD} = 0.070 \pm 0.009\, pe/initial\ e.$$

This is an order of magnitude lower compared to the yield of 0.71 pe/initial e obtained in gaseous Ar at 87 K at a gain of 400 [10]. Taking into account the average solid angle $\Delta\Omega/4\pi=2.7\times10^{-3}$, we obtained the following avalanche scintillation photoelectron-yield in Xe extrapolated to 4π acceptance:

$$Y_{PE} = 0.075 \pm 0.02\, pe/avalanche\ e.$$

Furthermore, accounting for the G-APD's photon detection efficiency averaged over the Xe emission spectrum of Fig. 3, <PDE>=10.7%, the avalanche scintillation photon yield over 4π can be estimated:

$$Y_{PH} = 0.70 \pm 0.2\, photon/avalanche\ e.$$

This photon yield in Xe is ~5-fold lower compared to the 3.8 photon/initial e obtained in Ar [10]. Note that the latter is ~15% smaller than that originally presented in [10], due to a revised calculation of the average PDE, amounting at 17.6% (averaged over the Ar emission spectrum presented in Fig. 3).

The THGEM/G-APD multiplier yields and avalanche scintillation yields in gaseous Xe (this work) and Ar [10] are summarized in Table 2.

| Medium | Gaseous Xe at 200K and 0.73atm | Gaseous Ar at 87K and 1.0atm |
|---|---|---|
| Double-THGEM charge gain | 350 | 400 |
| Initial (prior to multiplication) charge, electrons | 1700 | 900 |
| THGEM/G-APD yield, photoelectrons | 120 | 640 |
| THGEM/G-APD yield, photoelectrons per initial electron (prior to multiplication electron) | 0.070±0.009 | 0.71±0.09 |
| Avalanche scintillation yield over 4π, photoelectrons per avalanche electron | 0.075±0.02 | 0.66±0.2 |
| Avalanche scintillation yield over 4π, NIR photons per avalanche electron | 0.70±0.2 | 3.8±1 |

Table 2. A summary of THGEM/G-APD photoelectron yields and avalanche scintillation-photon yields, measured here in gaseous Xe at 200 K and 0.73 atm and that in gaseous Ar in the two-phase mode at 87 K and 1.0 atm [10]. $^{241}$Am X-rays converted in gaseous Xe and liquid Ar; G-APD viewing angle: ±70º. The data were corrected for G-APD nonlinearity and cross-talk.

The general conclusion of this section is that the THGEM/G-APD multiplier photoelectron yield in Xe, normalized to a full 4π solid angle, is about an order of magnitude lower than that in Ar; it is in accordance to the overlap of the G-APD's PDE spectrum with that of the noble-gas emission in the NIR range (limited to 950 nm) (Fig. 3). The avalanche-scintillation photon yield in Xe, normalized to 4π, in this NIR range is estimated to be about 5-fold lower than in Ar.

Due to these facts, Xe-based detectors with optical readout using bare G-APDs should be considered inapplicable in rare-event experiments requiring single-electron sensitivity, in contrast to Ar-based detectors. On the other hand, the expected THGEM/G-APD yield, of the order of 1000 photoelectrons per 511 keV gamma ray as deduced from table 2, is sufficient for PET applications. In addition, with Xe, better results could be expected with WLS-coated G-APDs, sensitive to its more copious VUV emission, as discussed in [44]. An alternative solution for THGEM optical readout in Xe might be InGaAs photodiodes of high sensitivity up to 1700 nm [48], i.e. within Xe major emission range.

6. **Conclusions**

The performances of THGEM multipliers have been studied for the first time in two-phase Xe, in view of potential applications in rare-event experiments and PET. Additional comparative measurements were performed in gaseous Xe, under cryogenic conditions, including investigations of optical recording of the avalanche with a G-APD.

The present work concludes a series of studies of GEM and THGEM multipliers in two-phase noble-gas avalanche detectors; we therefore summarized the available data accumulated with these hole-multipliers in two-phase Ar, Kr and Xe detectors.

Stable operation of a double-THGEM multiplier, of 2.5×2.5 cm$^2$ active area, was demonstrated in two-phase Xe; charge gains reached values of 600 at rather low noise rate, of the order of few Hz. The maximum gain attained, limited by discharges, was about 5-fold lower than in two-phase Ar. Comparative studies in gaseous Xe indicated that the maximum gain and the operation voltage are determined essentially by the gas density; i.e. any eventual Xe vapour condensation had no influence on the THGEM performance in two-phase Xe.

The optical readout of THGEM multipliers with G-APD was further investigated here in gaseous Xe at 200 K, recording avalanche scintillations in the NIR at wavelengths limited to 950 nm. At avalanche charge gain of 350 and under ±70º G-APD viewing-angle, the double-THGEM/G-APD multiplier yielded 0.07 photoelectrons per initial electron (prior to multiplication). This is an order of magnitude lower yield compared to that reached in gaseous Ar in the two-phase mode at similar gain value [10]; it was shown that the lower yield is consistent with the lower overlap of the G-APD's PDE and the Xe emission spectrum. The avalanche scintillation yield in Xe over the present NIR range, amounted at 0.7 NIR photons per avalanche electron, over 4π; it is 5-fold lower than that measured in Ar.

Both avalanche charge gains and avalanche scintillation yields obtained in the present work in two-phase and gaseous Xe could be sufficient for PET applications. On the other hand, unlike two-phase Ar detectors, the yields reached in Xe are not sufficient for conceiving detectors with ultimate sensitivity required for rare-event experiments. Alternatively, the optical readout with either WLS-coated G-APDs, sensitive to VUV emission of Xe, or with InGaAs photodiodes, sensitive to major NIR emission of Xe, should be investigated, provided the latter could be economically produced.

## 7. Acknowledgements

We are grateful to Y. Tikhonov for the support, R. Snopkov and A. Chegodaev for the development of the experimental setup. This work was supported in part by RFFI grant 09-02-12217-ofi_m, by special Federal Program "Scientific and scientific-pedagogical personnel of innovative Russia" in 2009-2013 and by The Israel Science Foundation grant 477/10. A. Breskin is the W.P. Reuther Professor of Research in the Peaceful use of Atomic Energy.### References